\documentclass[12pt]{iopart}

\usepackage{graphicx}
\usepackage{dcolumn}
\usepackage{bm}
\usepackage{color}
\usepackage{calrsfs}

\begin{document}

\title{GHz photon-activated hopping between localized states in a silicon quantum dot}

\author{T~Ferrus$^{1, \star}$, A~Rossi$^1$, A~Andreev$^1$, T~Kodera$^{2, 3, 4}$, T~Kambara$^{2}$, W~Lin$^{2}$, S~Oda$^2$ and D~A~Williams$^1$}
\ead{taf25@cam.ac.uk}
\address{$^1$ Hitachi Cambridge Laboratory, J. J. Thomson Avenue, CB3 0HE, Cambridge, United Kingdom}
\address{$^2$ Quantum Nanoelectronics Research Centre, Tokyo Institute of Technology, 2-12-1 Ookayama, Meguro-ku, Tokyo, 152-8552 Japan}
\address{$^3$ Institute for Nano Quantum Information Electronics, University of Tokyo, 4-6-1, Komaba, Meguro, Tokyo, Japan}
\address{$^4$ PRESTO, Japan Science and Technology Agency (JST), Kawaguchi, Saitama 332-0012, Japan}

\begin{abstract}

We discuss the effects of gigahertz photon irradiation on a degenerately phosphorous-doped silicon quantum dot, in particular, the creation of voltage offsets on gate leads and the tunneling of one or two electrons via Coulomb blockade lifting at 4.2\,K. A semi-analytical model is derived that explains the main features observed experimentally. Ultimately both effects may provide an efficient way to optically control and operate electrically isolated structures by microwave pulses. In quantum computing architectures, these results may lead to the use of microwave multiplexing to manipulate quantum states in a multi-qubit configuration.

\end{abstract}

\maketitle

\section{Introduction}

Achieving successful control of the tunneling of an electron through the barrier of a nanoscale device is a major step towards the ability to perform reliable operations and sensitive detection. 

Such a control is indeed required for industrial level manufacturing. In spin-related applications and, in particular, quantum computing one should be able to spin-filter the signal via Pauli blockade \cite{Pauli blockade} and achieve a high fidelity spin to charge conversion \cite {spin qubit} without compromising on the coherence time due to feedback actions from the detector \cite{Morello}. Such a requirement on the barrier properties applies to charge qubit implementations where the tunneling of a single electron realizes the unitary NOT operation \cite{Ferrus NJP, Gorman}. Again, in quantum cellular automata \cite{QCA}, the ultimate performance of the structure is determined by the tunneling time through the barrier. Therefore, one needs to be able to control the quantum dot tunnel barriers both statically and dynamically. 

For the former, barriers can be shaped by creating constrictions at the process stage \cite{doped SET} or by manipulating directly the electrostatic potential \cite{MOS SET}. In the first case, the presence of an interface with materials of different permittivity induces a depletion region that raises the electrostatic potential preferentially at the location of the constriction. Such a technique is mostly used in silicon devices. The tunneling probability is then determined by the doping, the shape of the constriction and eventually weak localization effects \cite{Ferrus AIP}. Tunnel barriers can also be created electrostatically by the use of split-gates \cite{split gate} in III-V materials or by forming a metal-oxide-semiconductor layer in silicon devices \cite{Angus}. The barrier transparency and shape are then adjusted by the gate voltage for a given gate width. This provides better controllability but creates an effective path for high frequency noise to propagate to the device and renders scalability complex.

To achieve dynamic control, DC voltage pulses are generally applied to the metal gates so that tunnel barrier height can be modulated allowing electrons to tunnel in or out of the dot. This technique has been successfully implemented on single electron transistors when used as electron pumps \cite{pump}, including turnstiles \cite{turnstile} and ratchets \cite{ratchet}. Despite its high efficiency the method still requires high frequency lines very close to the device.

An alternative method consists in sending photons at the matching frequency $h\nu = E_0-E_{\textup{\tiny{F}}}$ between the Fermi level $E_{\textup{\tiny{F}}}$ of the lead and a predefined level $E_0$ in the dot, thus opening a new channel for transport with a significant inelastic contribution. This is the principle of photon assisted tunneling (PAT) \cite{PAT}. Because the photon flux can be modulated, manipulation can be performed and the need for high frequency lines can be avoided. However, such a process to be effective has to be bound to conditions on the frequency range ($k_{\textup{\tiny{B}}}T\ll h\nu $) due to the discrete nature of the photon, as well as on the tunneling rate $h\it{\Gamma} <h\nu $ (non-adiabacity). Outside these conditions, PAT is not observable and the classical picture prevails, the device operating as a pump as described before.

Such a optical control using PAT is a well adapted method for Metal-Oxide-Semiconductor structure where the tunneling levels inside the dot can be well defined by confinement. In a doped material, randomness in dopant distribution both in space and in energy, induce the formation of an impurity band that progressively merges with the conduction band when increasing the doping concentration. This description is accurate in mesoscopic devices. However, in nanostructures of less than 100 nm typically, interface effects enhance localization effects due to dielectric screening, even in highly doped material. The impurity band may then become wider than $h\nu $ and the usual classical and adiabatic description of tunneling is expected to prevail on PAT, even at low temperatures. Consequently one cannot expect PAT to be present in doped-silicon quantum dots at liquid helium temperature and, apart from electron heating from inelastic processes and rectification in the leads, there should not be any significant effects from photon irradiation.

Nevertheless, if the structure of the quantum dot allows the internal dynamic to occur at a rate slower than the tunneling rate and/or if the electron-phonon coupling is weak, then non-adiabaticity can be recovered. Furthermore, if electron interaction is sufficiently strong, charging energy can be reduced effectively and electrons can tunnel \cite{Shklovskii}.

In this article, we demonstrate that under the condition $h\it{\Gamma} <h\nu \ll k_{\textup{\tiny{B}}}T$, non-adiabatic tunneling of single electrons into a highly doped quantum dot can be induced by microwave absorption and despite Coulomb repulsion only due to the presence of both localized states and electron-electron interaction. This regime differs significantly from both PAT and the one of electron pumps.

After a short introduction to the device structure and the experimental setup, we discuss already-known microwave effects that give resonances in the gigahertz (GHz) regime : cavity resonances in section \ref{CR}, rectification in section \ref{REC} as well as photon assisted tunneling in section \ref{PET} and spatial Rabi oscillation in section \ref{SRO}. We then describe the appearance a new type of resonances in the same range but in doped devices, at which Coulomb oscillations shift in gate voltage (Sec. \ref{Shifts}). In sections \ref{Indirect} and \ref{Direct}, we discuss the physical processes occurring respectively at the leads and at the quantum dot. Section \ref{Power} describes simulations of both the frequency and the power dependences of the current using experimental data. The question of microwave heating and specific phenomena happening at high power are discussed in Section \ref{Heating} before concluding the paper. Readers interested in internal dynamics and feedback actions on the tunneling electrons may refer to the Appendix, on single shot measurements.

\section{Devices and measurement setup} \label{Setup}

Two types of devices were used in these experiments. Both were fabricated from a 30 nm-thick silicon-on-insulator (SOI) wafer (Inset fig. 1b). In a first set of devices, phosphorous was implanted through a thin protective silicon oxide to provide a dopant density of $\sim 2\times 10^{19}$\,cm$^{-3}$ after thermal annealing. A single dot of diameter $\sim$75\,nm diameter as well as a side gate, the source and the drain contacts were patterned by electron beam lithography and reactive ion etching. Defect passivation was realized by growing a 15\,nm thermal oxide. In a second set of devices, the intrinsic silicon was dry etched to pattern the dot, the gate and the contacts, before depositing a 125\,nm of SiO$_2$ and subsequently a 200\,nm poly-silicon gate that covers the entire device region. In this configuration, a positive top-gate voltage induces a two-dimensional electron gas at the Si-SiO$_2$ interface. This allows the source, the drain and the underlying-gate leads to operate and the quantum dot to be populated by electrons.

Measurement were performed by a custom low temperature complementary metal-oxide-semiconductor circuit (LTCMOS) that provides the voltages to the gates and measures the SET current $I_{\textup{\tiny{SD}}}$ through a charge integrator. This arrangement and its advantages have already been described elsewhere \cite{Ferrus JAP}. All lines were filtered by single stage low-pass resistance-inductance-capacitor filters with a cut-off of about 80\,kHz to suppress electrical noise and minimize the electron heating. Devices and the LTCMOS measurement circuit were immersed into liquid helium at 4.2\,K. Microwave coupling was achieved by realizing a two-turn coil in close proximity of the device that was impedance-matched and connected to an Agilent E8257D-520 PSG Analog Signal Generator (PSG) via a high frequency line. Microwave pulsing experiments were realized by using the internal pulse pattern generator of the PSG with either an internal triggering for continuous pulse measurement or via the LTCMOS trigger for single shot measurements.

\section{Microwave spectrometry and microwave effects}

Microwaves affect semiconductor nanostructures in many ways for example by promoting cotunneling \cite{Manscher} or electron localization \cite{microwave}. This provides an elegant way of accessing information on devices characteristics or transport phenomena. In particular, microwave spectroscopy constitutes a well adapted method of investigation of the energy level structure in nanoscale objects \cite{Prati} and allows studying electron dynamics. We will first describe some of these effects in the context of quantum dots before describing new effects specific to similar but doped devices.

\subsection{General effects in quantum dots}

\subsubsection{Cavity resonance and heating\\}\label{CR}

Cavity resonances are intrinsic to experimental setups and result from reflections at each end of the line carrying the fast signal. These are generally observable in the radio-frequency (RF) range due to typical cable lengths, but harmonics could be generated up to several GHz. These modify the effective RF or microwave power at the device, and so modify electron cotunneling \cite{Manscher} and eventually electron heating at high power. If the device is a quantum dot, Coulomb oscillations are barely affected with the exception of an increased background conductivity, a lower visibility and eventually a larger Coulomb peak width. Ultimately, resonances disappear at high frequency due to signal attenuation. As a result, the resonance position in frequency is independent on both the gate voltage $V_{\textup{\tiny{g}}}$ and the source-drain bias $V_{\textup{\tiny{SD}}}$ despite some dependences on the resonance height and width due to cotunneling effects (Fig. 1).

\subsubsection{Rectification\\}\label{REC}

Upon irradiation by waves at frequency $\omega$, the effective bias applied to a quantum dot acquire an oscillatory term so that $V_{\textup{\tiny{SD}}}=V_{\textup{\tiny{dc}}}+\alpha  V_{\textup{\tiny{ac}}} \textup{sin} \left(\omega t+\phi \right)$ where $\alpha$ is a coupling constant. Due to the non-linearity in the current-bias dependence, the averaged measured current becomes $I_{\textup{\tiny{SD}}}\approx I_{\textup{\tiny{dc}}}+\beta V_{\textup{\tiny{ac}}}^2$ at small driving amplitudes. Consequently, the application of an ac-signal results in an increase in the source to drain current and leads to a modification of the source and drain capacitances without modifying the dot charging energy. This is purely a classical effect and no variation in the position of the resonance frequency in gate voltage is expected (Fig. 1). Other phenomena may lead to rectification as an indirect effect when the current-bias dependence is asymmetric \cite{Rectification}. Some of these effects are described in the next sections.

\subsubsection{Photon assisted tunneling and electron pumping\\}\label{PET}

Photon assisted tunneling is an inelastic process in which a photon of energy $h\nu$ allows electrons to tunnel into higher energy states \cite{PAT}. Because of the discreteness of the photon character, such process only applies at high frequency $\nu\gg 4k_{\textup{\tiny{B}}} T/h$ and so, above 350 GHz at 4.2 K. It is characterized by the formation of Coulomb peak sidebands at $\pm n h\nu$ if considering a $n$-photon process and may lead to a significant pump current in case of barrier asymmetry. At fixed $V_{\textup{\tiny{g}}}$ and $V_{\textup{\tiny{SD}}}$, there is an infinite number of resonances in the frequency spectrum corresponding to electron transitions between the Fermi level of one of the leads and a level in the quantum dot. Because quantized energy levels can be tuned by the gate voltage, all resonances have their positions in frequency varying linearly with $V_{\textup{\tiny{g}}}$.

At frequencies $\nu\ll 4k_{\textup{\tiny{B}}} T/h$, generally in the MHz range \cite{pump frequency}, the adiabatic picture prevails. Devices can be used as electron pumps or turnstiles \cite{pump} by modulating directly the height of tunnel barriers and so allowing electrons to tunnel in and out of the dot. The operation frequency can then be adjusted so that only one electron per cycle can tunnel. This leads to the formation of plateaus in the current-voltage characteristics at multiples of $e\nu$. In this case, the pumped current varies linearly with frequency and there is no specific resonance associated with the process.

\subsubsection{Spatial Rabi oscillations\\}\label{SRO}

When two spatially separated non-degenerate electronic orbitals in electron reservoirs are connected via a tunnel barrier like in a double quantum dot, the absorption of photons via PAT can induce charge transfer between the two dots and spatial coherence of the electronic states can occur. This process is known as spatial Rabi oscillations (SRO) \cite{Stafford}. Conditions for its observability are similar to the ones for PAT, and so the frequency spectrum is composed of a series of $N$ resonances at $h\nu = \left( \Delta \epsilon^2 + 4 w^2 \right)^{1/2}/N$ where $\Delta \epsilon$ is the difference detuning energy between the two dots and $w$ is the tunneling matrix element between the dot orbitals.

Photon-induced spatial Rabi oscillations have also been reported in undoped silicon metal-oxide-semiconductor (MOS) structures \cite{hasko solgel, Ferrus unpublished} as well as in single dot structures \cite{hasko doped}. More recently similar experiments have been performed in silicon nanowires containing two dopants only \cite{CEA}. In these later cases, the effect is linked to the presence of localized states whether they are due to Pb centers at an silicon-oxide interface \cite{PB} or due to implanted donors like Phosphorous or Arsenic. Microwaves provide energy difference between the levels of two distant traps as long as screening by the other surrounding electrons is negligible. 
Such energies are reflected in the microwave dependence of the current by small and sharp resonances, generally a few tens of kilohertz-wide. The shape of these resonances depends on the position of the two trapping sites relatively to the current path, so that larger but wider resonances always correspond to traps that are located close to the tunnel barriers and interact strongly with conducting electrons, leading to a short $T_1$ and $T_2$ lifetimes with $T_1\gg T_2$. Away from the conducting path $T_2\sim 2 T_1$ could be obtained due to inelastic processes being dominant. 
Because of the long $T_1$ lifetime in doped quantum dots due to the poor electron-phonon coupling \cite{E-P coupling, Hasko3}, electrons can be transferred coherently from one site to the other despite $h \nu\ll k_{\textup{\tiny{B}}} T$ \cite{Erfani}, the relevant parameter becoming the tunneling time rather than the temperature. 
The position of these resonances in the frequency spectrum depends on the electrostatic potential that modifies the energy levels and so, they are strongly dependent on both the source-drain bias $V_{\textup{\tiny{SD}}}$ and the SET gate $V_{\textup{\tiny{g}}}$.

\subsection{Doped quantum dots under microwave irradiation}

Megahertz or GHz signals are generally applied via fast lines to metal gates in proximity of the device. In our arrangement, coupling between the microwave signal and the device is realized by a small but fast oscillating magnetic field oriented perpendicular to the SET, avoiding the need for a strip-line \cite{coupling}. Despite strong signal attenuation at the center of the coil, this method allows matching the circuit impedance regardless of the nature of the device and authorizes contactless manipulations. Still, when the latter method is used, it is necessary to distinguish between microwave induced-electron excitations in the leads, including the SET gate, the source and the drain contacts, that \textit{indirectly} affect the dot conductivity, and \textit{direct} excitations of electrons across the dot tunnel barriers of the quantum dot. These are described in the following sections.

\subsubsection{Coulomb peak shifts\\}\label{Shifts}

\begin{figure}
\begin{center}
\includegraphics[width=85mm, bb=0 0 331 468]{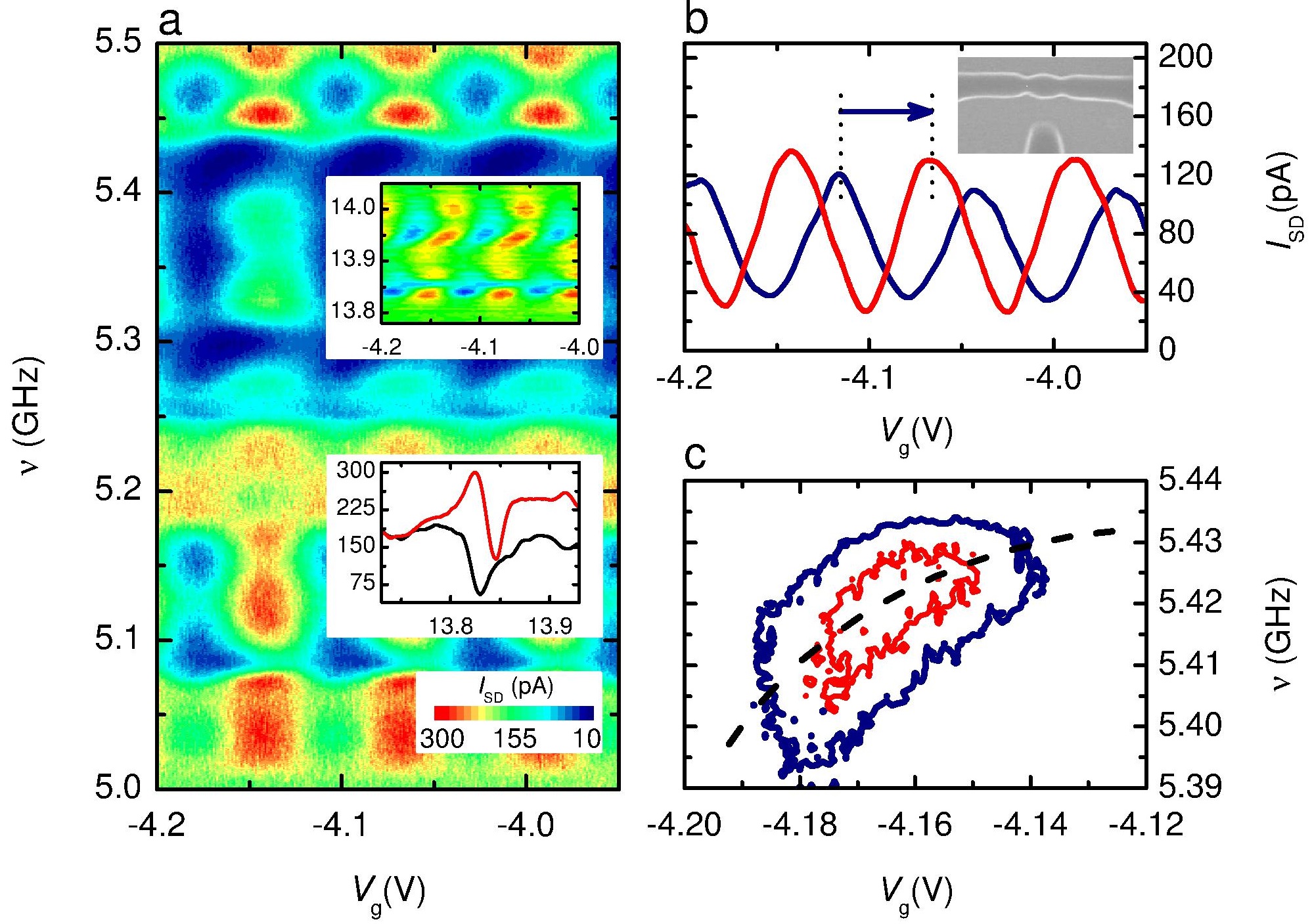}
\end{center}
\caption{\label{fig:figure1} (a) Coulomb oscillations as a function of frequency at fixed microwave power $P=$-6 dBm. Bleaching effects where either levels are involved in the transport are observed at around $\nu \sim 5.2$ GHz \cite{bleaching}. Top inset : similar dependence around 14 GHz. Bottom inset : corresponding frequency resonances on top (red) and at the bottom (black) of a Coulomb peak. (b) Coulomb oscillations at $\nu=5.435$ Ghz and $P=-6$ dBm (red) and, out of resonance at $\nu=5.384$ Ghz and $P=-6$ dBm (blue). The inset shows an SEM image of the device with source, drain and gate leads. (c) Constant current contour plots around $\nu=5.4$ Ghz. The dashed line shows the dependence of the resonance frequency on gate voltage.}
\end{figure}

\begin{figure}
\begin{center}
\includegraphics[width=85mm, bb=0 0 331 468]{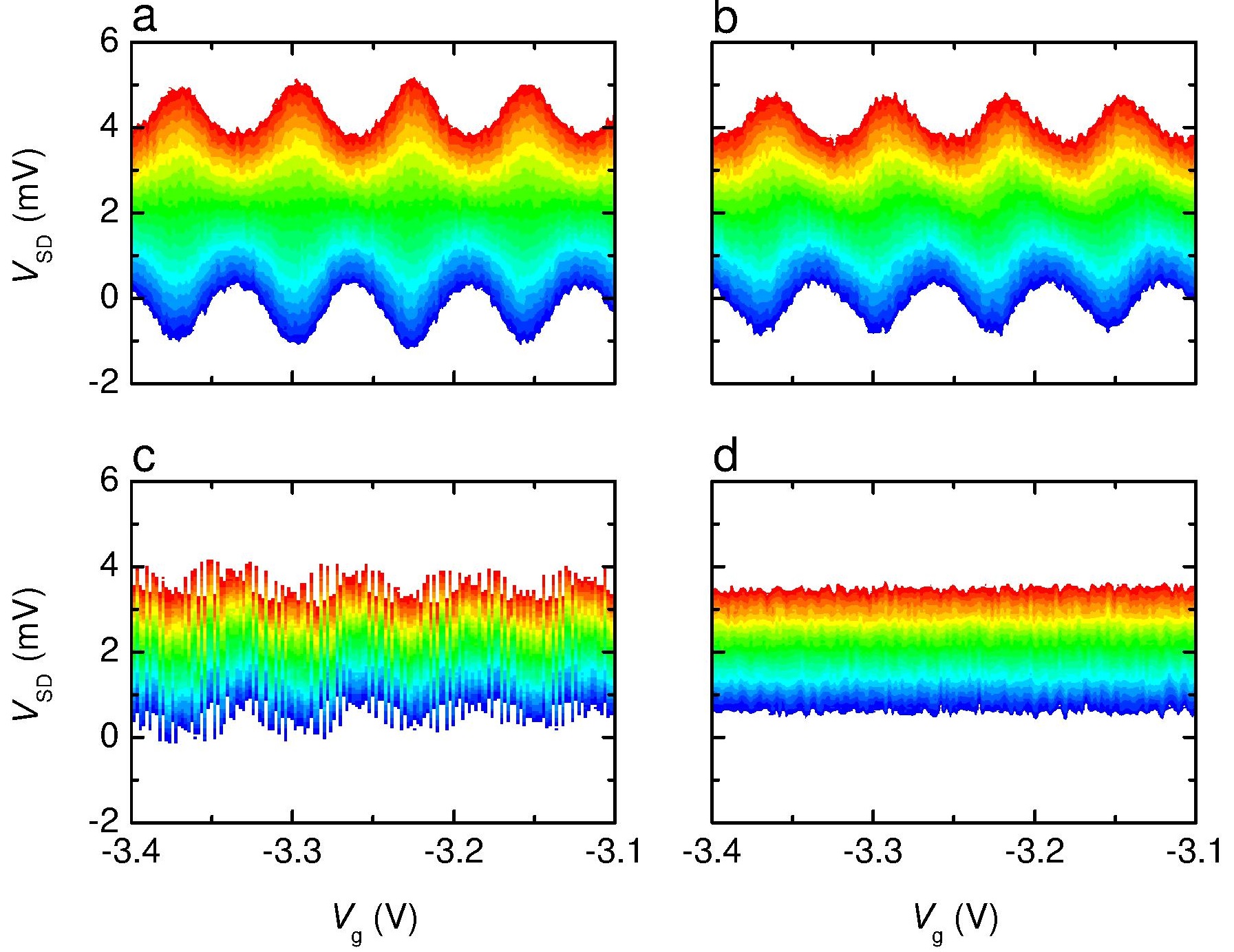}
\end{center}
\caption{\label{fig:figure2} Contour-plot of $I_{\textup{\tiny{SD}}} \left( V_{\textup{\tiny{g}}}, V_{\textup{\tiny{SD}}}\right)$ for device A at fixed frequency $\nu=12.2$ GHz and microwave power $P=$-135 dBm (a, undisturbed), -3 dBm (b), 6 dBm (c), 9 dBm (d).}
\end{figure}

\begin{figure}
\begin{center}
\includegraphics[width=85mm, bb=0 0 331 468]{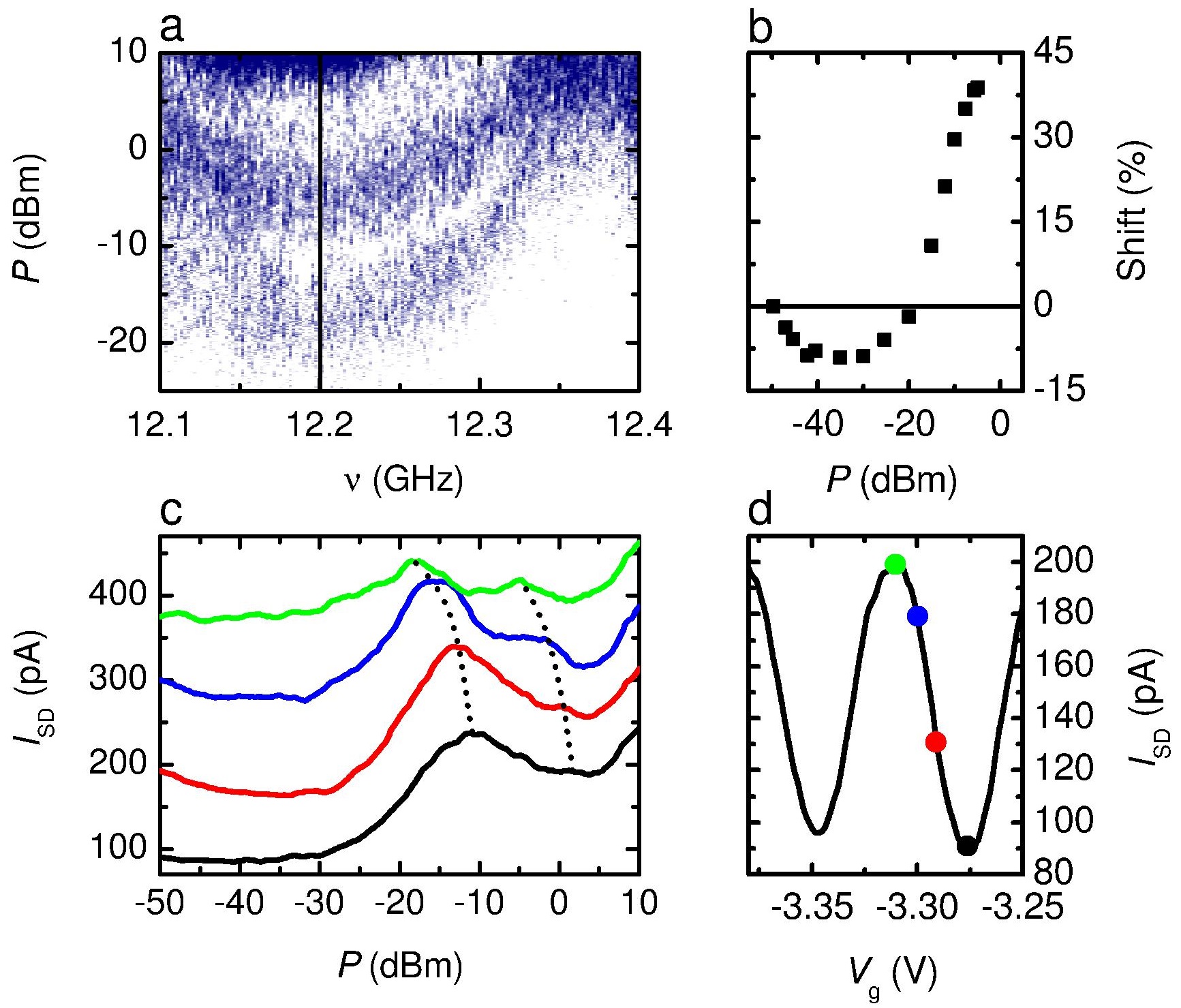}
\end{center}
\caption{\label{fig:figure3} (a) Variation of $I_{\textup{\tiny{SD}}}$ with $P$ and $\nu$. (b) Power dependence of the shift at $\nu = $12.2 GHz, in unit of the Coulomb blockade oscillation period. The horizontal line indicates the 50 $\%$ limit.(c) Power dependences of $I_{\textup{\tiny{SD}}}$ at $\nu=12.2$ Ghz taken at different value of $V_{\textup{\tiny{g}}}$ as marked in (d). Curves are offset for clarity. Source-drain bias was fixed at $V_{\textup{\tiny{SD}}}=3.5$ mV.}
\end{figure}

In doped devices, the dependences of the SET current $I_{\textup{\tiny{SD}}}$ on gate voltage $V_{\textup{\tiny{g}}}$ and microwave frequency $\nu$ reveal a singular behavior with Coulomb peaks shifting along $V_{\textup{\tiny{g}}}$ at specific frequency values (Figs. 1a and b). At fixed $V_{\textup{\tiny{g}}}$ and $V_{\textup{\tiny{SD}}}$, the conductivity then acquires a dependence on frequency which leads to the appearance of resonances in the frequency spectrum. Their positions depend on $V_{\textup{\tiny{g}}}$ as attested by the asymmetry in the iso-current contours (tear-like shape) around the resonances (for example at $\nu \sim 5.1, 5.3$ or 5.4 GHz) (Fig. 1c). Their height is also within the range of the Coulomb peak height for a given microwave power and the extent of the shift is never larger than 0.5\,$e$. These only cannot be explained by cavity resonance, microwave heating \ref {CR} or rectification alone \ref{REC}. 

On the other hand, involving \textit{standard} SRO or PAT is equally difficult. Indeed conditions for PAT are not matched, especially on temperature and characteristics in frequency are different. Also, unlike our experiments, SROs observed in single dot structures \cite{hasko doped} did not involve shifts of Coulomb peaks in gate voltage as these concerned internal dynamics and so the dot charge was conserved at all time. The width of these new type of resonances is also generally broader (few 10 to few 100 MHz) compared with SRO's (few 10 kHz) (Fig. 1a). Because it depends on value of the shift as well as the shape of the Coulomb peak itself, these additional resonances do not give direct information on the energy relaxation time $T_1$ of the system \cite{T1, Ferrus JAP}.

Microwave-induced gate voltage shifts are observed for a wide range of frequencies up to 15 GHz where signal attenuation in the coaxial cable becomes significant. However, in order to get an insight into the process responsible for this effect, we will restrict the analysis to a couple of frequencies around 5 and 12 GHz. In particular, we chose a resonance that gave the maximum shift in gate voltage and we analyzed the current dependences on $V_{\textup{\tiny{g}}}$ and  $V_{\textup{\tiny{SD}}}$ at different microwave powers $P$ (Fig. 2). Four regimes are clearly visible : (i) at low power, there is a small shift in $V_{\textup{\tiny{SD}}}$ only (Fig. 2a), (ii) at moderate power, diamonds start getting distorted and shift mostly towards $V_{\textup{\tiny{g}}}>0$ if  $V_{\textup{\tiny{SD}}}>0$ and towards $V_{\textup{\tiny{g}}}<0$ if $V_{\textup{\tiny{SD}}}<0$ leading to a sawtooth like behavior (Fig. 2b), (iii) at moderately high power, diamonds are a superposition of the diamonds in the absence of microwaves and the diamonds shifted (Fig. 2c), (iv) at high power, rectification in the leads becomes non-negligible and the diamonds decrease in size. Further increase in power leads to the disappearance of diamonds despite the S-shape dependence of the current in $V_{\textup{\tiny{SD}}}$ still being clearly visible (Fig. 2d). This is not due to heating but to higher order tunneling induced at high power (see Sec. \ref{Heating}).

Such a change with microwave power $P$ indicates that a complex process is taking place and brings additional arguments against \textit{standard} PAT model. A more detailed study shows that the shift does not have a monotonic variation with $P$ and that its sign changes indicating the presence of two distinct processes, one at low power and a second that becomes active above $\sim -30$ dBm (Fig. 3b). Saturation is expected at high power but not observed due to charge configuration instabilities in the dot (Fig. 2c and Sec. \ref{Heating}). Nevertheless the most striking result is the presence of a double-peak feature in the power dependence of the current. Both its visibility and position depend on frequency (Fig. 3a) and gate voltage (Figs. 3c and d). Such a feature, on its own, is distinct from PAT's (see Sec. \ref{PET}).

Finally, we should highlight that these effects are observed in all doped devices independently of the fabrication process, but do not occur in undoped devices under normal operation (structure in the inversion mode). However, under the conditions of quasi-depletion on the leads (low electron density), shifts of much smaller amplitude can be obtained ($\sim$ mV). By depleting electrons from the leads, the influence of localized states on conductivity, especially those at the Si-SiO$_2$ interface, is enhanced. This suggests the involvement of trap sites in the explanation for the gate voltage shifts, either from defects or surface states in undoped devices or from the phosphorous dopants in doped ones.

Individual process steps (doping, etching, annealing and oxidation) being executed indistinctly to all sections of the device at the same time (dot and leads), the different sections of the device are expected to be structurally similar. However, due to the various dimensions in the device sections (from $\sim$ 30 nm for the dot region to few $\mu$m in the leads), the electrostatic potential and so the localization strength is significantly different across the device. Because the microwave is being broadcast to the whole device, it is necessary to distinguish induced effects in the dot, the gate lead and, the source and drain contacts.

\subsubsection{Indirect microwave effects : voltage drop-off and rectification in the leads\\}\label{Indirect}

Despite the high dopant concentration, a region of localization does exist near the Si-SiO$_2$ interface because of the weakening of the electron screening effect, i.e. increase in the strength of long-range interaction, due to dielectric screening \cite{dielectric screening} (Fig. 4b). The boundary between extended and localized states in then defined by an ionization energy for the phosphorus dopants $E_{\textup{\tiny{B}}} \sim k_{\textup{\tiny{B}}} T$. This localization always leads to a voltage drop-off $\mid V_{\textup{\tiny{g}}}-V_{\textup{\tiny{g}}}^* \mid$ at the end of the gate. Because materials in the gate leads, drain and source are identical and processed identically, microwave-induced SROs are also active in these regions. If the donor pair is away from the Si-SiO$_2$ interface, the effect is expected to be electrostatically screened by the other surrounding electrons because of the high doping concentration, a situation which is similar to the bulk case \cite{screening}. However, if the pair is in proximity of the interface, the situation is more complex (Fig. 4a). Electrostatic symmetry breaks up and SRO between pairs located perpendicular to the interface give the strongest effect by dynamically allowing charges to rearrange and so to minimize the drop-off voltage. This sets a new effective voltage $\mid V_{\textup{\tiny{g}}}^* \left( \nu \right)\mid < \mid V_{\textup{\tiny{g}}}^* \left( 0 \right)\mid$ at the end of the gate lead. Such effect is expected to be gate voltage independent but sensitive to both the microwave power and the frequency. Power defines how efficient the SROs are whereas frequency probes both the density of states of the donors and their spatial distribution in a complex fashion. When broadcasting microwaves, many pairs will be involved. However, due to randomness in dopant distribution, the difference $\mid V_{\textup{\tiny{g}}}^* \left( \nu \right) - V_{\textup{\tiny{g}}}^* \left( 0 \right)\mid$ will be small but non-zero. As a result, Coulomb peak shifts towards negative $V_{\textup{\tiny{g}}}$ to reflect the change in the voltage drop-off at the end of the gate. Because such an effect is weak, it can only be observed at low power when other microwaves effects are inactive (Fig. 3b). Despite being present in both the source and drain leads, the effect there is much minimized due to the interface being perpendicular to the charge transport and increased scattering.

\subsubsection{Direct microwave effects : electron tunneling through the dot barrier\\}\label{Direct}

A similar effect to the one described previously can be used to explain the small shift in source-drain bias, at very low power. However, at higher power, dynamic effects dominate. The observed distortion of the current contour in Figure 2 is due to a modification of the tunnel conductances that become asymmetric with $V_{\textup{\tiny{SD}}}$ when the microwave power is increased. However, the shape of Coulomb diamonds determined by the contour-plots of $\textup{d} I_{\textup{\tiny{SD}}} / \textup{d} V_{\textup{\tiny{SD}}}$ barely changes with $P$. Only the switching behavior and the shift along $V_{\textup{\tiny{g}}}$ are observed. The latter can be interpreted as a change in the average electron number $\bar{n}$ in the quantum dot, with $0\leq \Delta\bar{n}\leq 1$ under the bias conditions for Coulomb blockade. Under microwave illumination one electron tunnels through the barrier, against Coulomb blockade at a rate $\it{\Gamma}_{h\nu}\sim {P}^{\gamma}$ where $P$ is the microwave power and $\gamma > 0$ (See Sec. \ref{Power}). Under Coulomb repulsion, the electron is tunneling back to the source or drain at a rate $\it{\Gamma}_{\textup{\tiny{S,D}}}$, depending on the bias conditions. In the absence of tunneling current through the dot (blockade regime), $\bar{n}$ is then modified so that

\begin{eqnarray}\label{eqn:equation1}
\Delta\bar{n} \sim \frac{\it{\Gamma}_{h\nu}}{\it{\Gamma}_{h\nu}+\it{\Gamma}_{\textup{\tiny{S,D}}}} 
\end{eqnarray}

Microwave-induced tunneling well explains the shift direction respectively to $V_{\textup{\tiny{SD}}}$ as the process tend to compensate the sense of natural tunneling and so the sign of current, set by $V_{\textup{\tiny{SD}}}$. $\Delta\bar{n}$ thus increases with $P$ and reaches the value $1/2$ once the induced tunneling process ${\it{\Gamma}}_{h\nu}$ compensates for the electron tunneling back to the contact to reestablishing blockade conditions. At higher power, more than one electron on average is transferred. Due the finite integration time for measurement ($\sim 12\,\mu$s), two alternating sets of diamonds are superimposed as observed in Fig. 2c (See Sec. \ref{Heating}). 
This process is distinct from the regular and commonly observed PAT. In particular, the electron population probability does not follow the ${J_0}^2\left( eV_{\textup{\tiny{ac}}}/h\nu\right)$ dependence \cite{PAT2} (Fig. 2) and the relationship between the microwave amplitude signal $V_{\textup{\tiny{ac}}}$ and the frequency $\nu$ seems to be quadratic (Fig. 3a).

As seen previously, such a process is only present in doped devices. In the absence of microwaves and in the blockade regime, electron tunneling between two donors with energies $E_i$ and $E_j$ located across a tunnel barrier is thermally and electrostatically prohibited because $E_{\textup{\tiny{C}}}\gg k_{\textup{\tiny{B}}}T> h\nu$. In doped devices and despite the high doping concentration, localized states exist in depleted or partially depleted regions of the device as demonstrated previously \cite{Ferrus AIP}. This includes the Si-SiO$_2$ interface but also the tunnel barriers. They are also present in Lifshitz band tails \cite{Lifshitz} in the source and drain reservoirs. In this case, localized states distant from each other by $r_{ij}$ can be found on both sides of the tunnel barrier that match the condition \cite{Efros}

\begin{eqnarray}\label{eqn:equation2}
E_j-E_i - e^2/r_{ij} =h\nu,
\end{eqnarray}

leading to electron tunneling from the source or drain lead into the SET island.

\begin{figure}
\begin{center}
\includegraphics[width=157mm, bb=0 0 1000 1400]{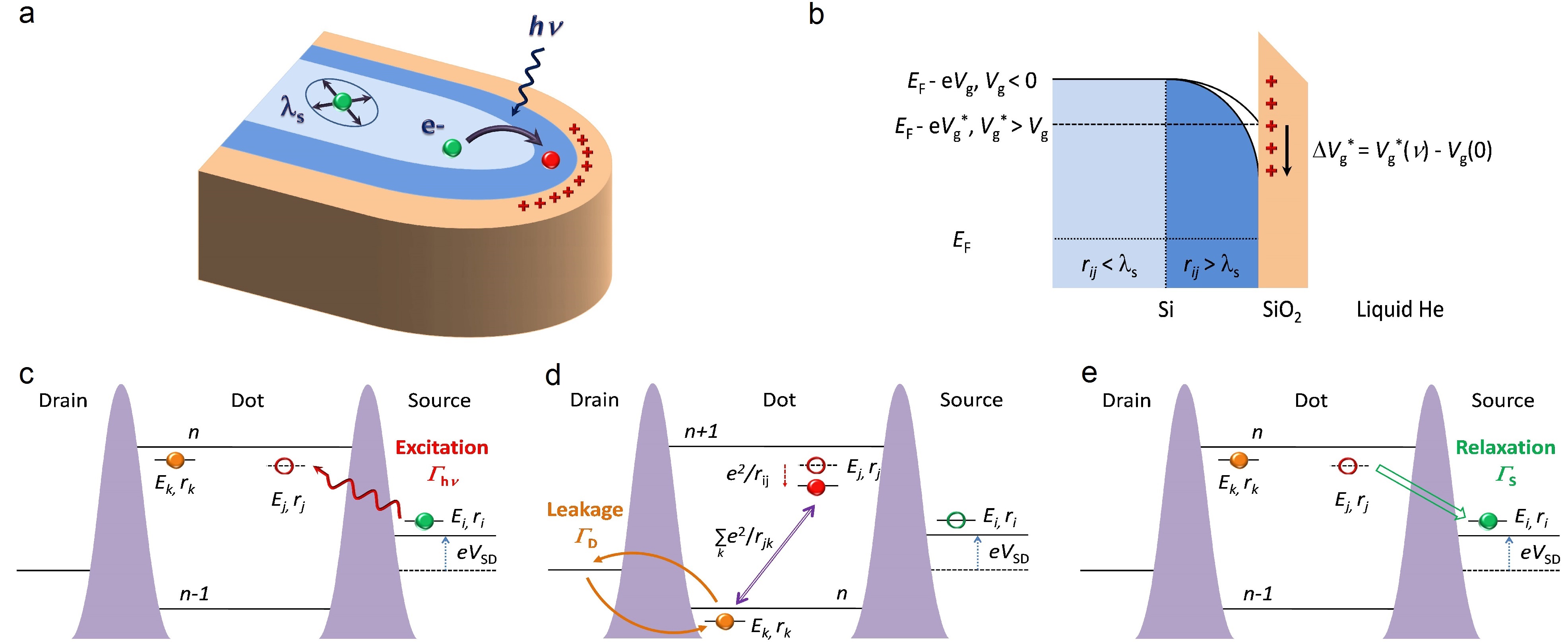}
\end{center}
\caption{\label{fig:figure4} (a) Delocalized-localized region in the gate lead with electron interaction screening effects. ${\lambda}_{\textup{\tiny{S}}}$ is the Thomas-Fermi screening length. (b) Change of effective potential in the lead in the case of transitions involving the localized region ($r_{ij} > \lambda_{\textup{\tiny{S}}}$). $E_{\textup{F}}$, $V_{\textup{\tiny{g}}}$ and $r_{ij}$ are respectively the Fermi energy, the gate voltage applied to the lead and the separation between the two localized states involved in the transition. (c)-(e) Details of the process involving the tunneling of a single electron using GHz photons : absorption of photons allows one electron to tunnel from the lead to a vacant site inside the dot (c); the tunneling electron interacts via Coulomb interaction with the other already present electrons inside the dot, inducing charge rearrangement and leading to an electron localized on a different trap to tunnel out into the drain lead (d); electrons are localizing back to their own traps (e). If $\it{\Gamma}_S <\it{\Gamma}_D$, there could be a short time interval during which the dot has only $n-1$ electrons.}
\end{figure}

Consequently, the excess of the charge in the island leads to Coulomb peaks shifting towards positive $V_{\textup{\tiny{g}}}$ above -30 dBm (Fig. 3c).

\section{Power dependence} \label{Power}

In this section, we describe qualitatively the change in current through the quantum dot as the microwave power is swept as a function of frequency (Fig. 3a) or gate voltage (Fig. 3c). If the current always monotonically increases with $P$ in undoped devices, the presence of a double peak in the power dependence of $I_{\textup{\tiny{SD}}}$ at specific frequencies (Fig. 3c) is a typical characteristic of doped devices. This results from the combination of various effects :

\begin{itemize}
  \item[(i)] a direct modification of the current $\Delta I_{\textup{\tiny{SD}}}$ due the microwave induced-electron tunneling in the quantum dot.

  \item[(ii)] a shift of the Coulomb peak position in gate voltage $U(P)$ as the microwave power is increased due to the excess effective charge in the dot and the modification of the voltage drop-off in the gate lead.
  
  \item[(iii)] rectification effects in the source, drain and gate leads due to the microwave capacitively coupling to the contacts.
  
  \item[(iv)] two-electron tunneling process at high power
   
  \item[(v)] electron heating via cotunneling at very high microwave power.
\end{itemize}

Simulating the results of figures 3a and 3c requires calculating the time averaged current $\overline{I_{\textup{\tiny{SD}}}}$. Indeed, in presence of a time-varying electromagnetic field, both $V_{\textup{\tiny{g}}}$ and $V_{\textup{\tiny{SD}}}$ acquire a time dependence and so does the instantaneous current $I_{\textup{\tiny{SD}}} \left( t \right)$. We have :

\begin{eqnarray}\label{eqn:equation3}
\overline{I_{\textup{\tiny{SD}}}} = \oint \limits_{0< \nu t <1} I_{\textup{\tiny{SD}}} \left[ V_{\textup{\tiny{g}}} (t), V_{\textup{\tiny{SD}}} (t)\right]\,dt
\end{eqnarray}

The integration path is here described by an ellipse in the $V_{\textup{\tiny{g}}}-V_{\textup{\tiny{SD}}}$ plane of figure 2a, defined by :

\begin{eqnarray}\label{eqn:equation4}
V_{\textup{\tiny{g}}}(t)  = V_{\textup{\tiny{g0}}} + \alpha\left(\nu \right) V_{\textup{\tiny{ac}}}\, \textup{sin} \left( 2 \pi \nu t \right) +U(P_{\textup{\tiny{eff}}}) \\
V_{\textup{\tiny{SD}}}(t) = V_{\textup{\tiny{SD0}}}+ \beta\left( \nu \right) V_{\textup{\tiny{ac}}}\, \textup{sin} \left( 2 \pi \nu t +\phi \right)
\end{eqnarray}

$V_{\textup{\tiny{g0}}}$ and $V_{\textup{\tiny{SD0}}}$ are the values of the gate voltage and the source-drain bias at which the measurement is taken. $V_{\textup{\tiny{ac}}} \propto \sqrt{P_{\textup{\tiny{eff}}}}$ is the microwave signal amplitude at the device and $P_{\textup{\tiny{eff}}}$ is the effective absorbed power. $P_{\textup{\tiny{eff}}}$ is generally different from the power $P$ at the output of the microwave source. This is not only due to attenuation but also depends on the variation of the absorption coefficient on the microwave intensity. Our experiments indicate that $P_{\textup{\tiny{eff}}} \propto \sqrt{P}$, a behavior that has already been observed in doped material including glasses where spectral diffusion, e.g. interaction with non-resonant thermally excited localized states via phonons, is absent \cite{Spectral diffusion}. This behavior is consistent with previous observations on similar devices \cite{E-P coupling}.

These equations do take into account the rectifying effects (iii) via the time integration. Here, $\phi$ represents the relative phase between the induced ac-signals. $\alpha$ and $\beta$ are the coupling strengths of the ac-signal to the gate and contact leads, respectively. Their dependence in frequency is well approximated by a Gaussian centered at the resonance frequency $\nu_{\textup{\tiny{0}}}$ (Inset in Fig. 1a) \cite{resonance}:

\begin{eqnarray}\label{eqn:equation5}
\alpha\left( \nu \right) = A\, {\textup{exp}} \left[ -\left(\nu-{\nu}_{\textup{\tiny{0}}}\right)^2/{W_{\textup{\tiny{g}}}}^2\right] \\
\beta \left( \nu \right) = B\, {\textup{exp}} \left[ -\left(\nu-{\nu}_{\textup{\tiny{0}}}\right)^2/{W_{\textup{\tiny{SD}}}}^2\right]
\end{eqnarray}

$A$, $B$, $\phi$ as well as $W_{\textup{\tiny{g}}}$ and $W_{\textup{\tiny{SD}}}$ are adjustable parameters. Finally, the function $U\left( P_{\textup{\tiny{eff}}} \right)$ represents the dependence of the shift value on $P_{\textup{\tiny{eff}}}$ (effect (ii)). Its values are given in figure 3b.

Calculation proceeds as follow. At a given time $t=t^*$ and power $P$, the effective gate voltage $V_{\textup{\tiny{g}}} (t^*)$ and the effective source-drain bias $V_{\textup{\tiny{SD}}}(t^*)$ are calculated and, the current $I_{\textup{\tiny{SD0}}}(t^*)$ can be extracted at these values directly from the experimental data in Fig. 2a. On the other hand, the tunnel rates $\it{\Gamma}_{\textup{\tiny{S}}} \sim \it{\Gamma}_{\textup{\tiny{D}}}$ can also be estimated experimentally by supposing symmetric tunnel barriers. The excess current $\Delta I_{\textup{\tiny{SD}}} (t^*)$ (i) can then determined by the change of the tunnel rate through the device due to electrons entering the dot :

\begin{eqnarray}\label{eqn:equation6}
\Delta I_{\textup{\tiny{SD}}} = e \frac{\it{\Gamma}_{h\nu}}{\it{\Gamma}_{h\nu}+\it{\Gamma}_{\textup{\tiny{D}}}+\it{\Gamma}_{\textup{\tiny{S}}}} \frac{{\it{\Gamma}_{\textup{\tiny{D}}}}^2}{\it{\Gamma}_{\textup{\tiny{D}}}+\it{\Gamma}_{\textup{\tiny{S}}}}
\end{eqnarray}

$\it{\Gamma}_{h\nu} = \gamma P_{\textup{\tiny{eff}}} \, \textup{exp} \left[ -\left(\nu-{\nu}_{\textup{\tiny{0}}}\right)^2/{W_{\nu}}^2\right]$ \cite{tunnel rate} is the additional tunnel rate induced by the microwave and $\gamma$ a constant.

The total current at $t^*$ is then

\begin{eqnarray}\label{eqn:equation7}
I_{\textup{\tiny{SD}}}(t^*) = I_{\textup{\tiny{SD0}}}(t^*) + \Delta I_{\textup{\tiny{SD}}}(t^*)
\end{eqnarray}

Repeating the process along the ellipse for $0<t<1/\nu$ allows to calculate the expected current $\overline{I_{\textup{\tiny{SD}}}}$ through the device at a given $P$ using equation 3.

Neglecting microwave induced electron tunneling ($U\left( P \right) = 0$ and $\Delta I_{\textup{\tiny{SD}}}=0$) only produces a monotonic variation of the current with $P$ (Fig. 5a) as expected. The presence of two peaks can only be obtained when $U\left( P \right) \neq 0$ (ii) whereas the peak positions and dependence on gate require (i) to be included in the equations (Fig. 5b). Optimizations are then performed so the positions of the peaks, their dependences in gate voltage and frequency, as well as the values of currents are best fitted. Results in figure 5 were obtained with ${\nu}_{\textup{\tiny{0}}} = 12.2$ GHz, $A \sim$ 0.009, $B \sim$ 0.0037, $\phi=0.55 \pi$, $\gamma =$ 17.5 nA.W$^{-1}$, $W_{\textup{\tiny{g}}} \sim W_{\textup{\tiny{SD}}} \sim$ 140 MHz and $W_{\nu} \sim$ 90 MHz. Values for $A$ and $B$ indicate the size of the ellipse is still smaller than the size of the Coulomb diamonds even at high power, thus discarding higher order tunneling effects. The first peak at $\sim -10$ dBm is well reproduced with a correct height, position in $P$ and dependence on $V_{\textup{\tiny{g}}}$. This results from the shift of the Coulomb peak when increasing $P$ whereas the second peak that appear more pronounced than in experiments, results from negative currents being compensated by $\Delta I_{\textup{\tiny{SD}}}$ when $V_{\textup{\tiny{SD}}} (t)<0$.

\begin{figure}
\begin{center}
\includegraphics[width=85mm, bb=0 0 331 468]{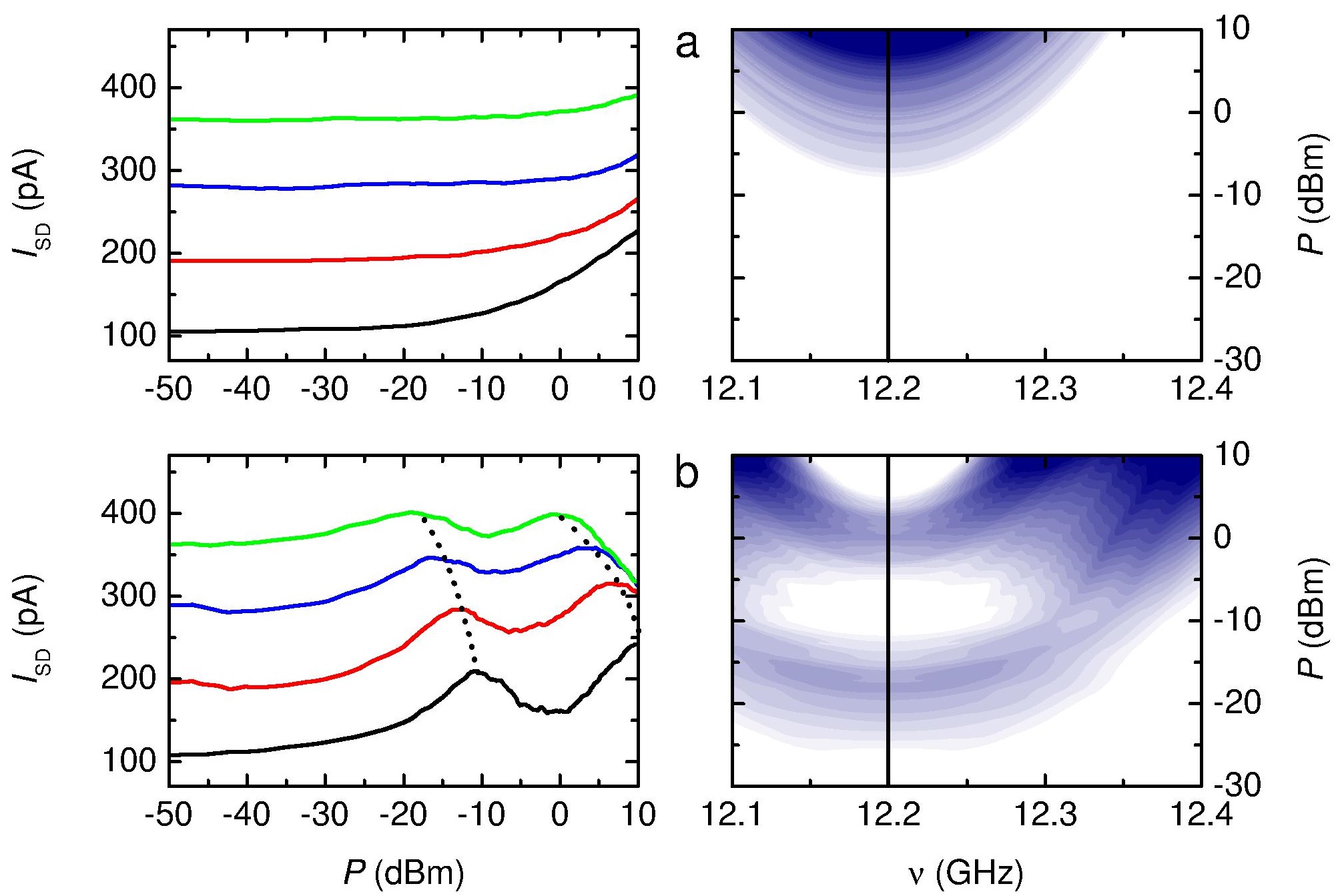}
\end{center}
\caption{\label{fig:figure5} Dependence of the dot current with microwave power at different gate voltages as set in Figure 3d (left) and dependences on power and frequency as measured in Figure 3a (right) in the case of, (a), an absence of microwave-induced tunneling and, (b), in presence of tunneling.}
\end{figure}

Microwave heating is negligible in our experiments and so, does not play a role even at high power. On the contrary, two-electron tunneling (iv) does attenuate rectifying effects due to microwaves (Sec. \ref{Heating}). However, the general characteristics, in particular the presence of the second peak, are not modified. Only its amplitude and position in $P$ are. Overall, it appears that the set of equations 3-9 well describe the microwave-induced dynamics in the quantum dot.

\section{Microwave heating and high power tunneling} \label{Heating}

One specific feature observed in the $I_{\textup{\tiny{SD}}} \left( V_{\textup{\tiny{g}}}, V_{\textup{\tiny{SD}}}\right)$ dependences is the presence of a bistable behavior along $V_{\textup{\tiny{g}}}$ for $P>$ 0 dBm (Fig. 2d). At fixed $V_{\textup{\tiny{SD}}}$, it appears as an alternation of two Coulomb oscillations, each one being shifted along $V_{\textup{\tiny{g}}}$ from their normal positions at $P = -135$ dBm by, respectively $\Delta V_{\textup{\tiny{g1}}}$ and $\Delta V_{\textup{\tiny{g2}}}$ (Fig. 6a). This effect is only obtained when acquiring the current dependence along $V_{\textup{\tiny{SD}}}$ while maintaining $V_{\textup{\tiny{g}}}$ at a fixed value. Due to the timescale of the switching phenomenon ($\sim \mu$s) being much smaller than the measurement time ($\sim$ms), all experiments in which $V_{\textup{\tiny{g}}}$ is swept do indeed measure an averaged current between the two oscillations. As a result the width of Coulomb peak appears to increase up to 6 dBm typically (Fig. 6b). At very high power, this averaging process leads to the disappearance of Coulomb oscillations (Fig. 2d). However, one can verify the effect is not induced by microwave heating by measuring the width at half height of the Coulomb peak from Figure 6a rather than by standard measurement of $I_{\textup{\tiny{SD}}} \left( V_{\textup{\tiny{g}}}\right)$ at fixed $P$ and $V_{\textup{\tiny{SD}}}$, and then deduce a value for the electron temperature $T_{\textup{\tiny{e}}}$. At 4.2 K, $k_{\textup{\tiny{B}}}T$ is much greater than the mean single particle energy spacing and so, tunneling occurs via multi-level and $e {\alpha}_{\textup{\tiny{g}}} W = 4.35 k_{\textup{\tiny{B}}} T_{\textup{\tiny{e}}}$ applies. Estimate values for $T_{\textup{\tiny{e}}}$ remains unaffected by an increase of $P$ (Fig. 6c), and microwave heating can be considered as irrelevant in our case.

On the other hand, variations of $\Delta V_{\textup{\tiny{g1}}}$ and $\Delta V_{\textup{\tiny{g2}}}$ in power calculated from Figure 6a shows a trend towards $\Delta V_{\textup{\tiny{g2}}} \sim 2\Delta V_{\textup{\tiny{g1}}}$ at very high $P$, suggesting that at such power a two-electron process is present (Fig. 6d). Indeed, by increasing $P$, the change in the average number of electrons in the dot $\Delta \bar{n}$ increases following Equation 1 until $\Delta \bar{n} \sim 1/2$. At higher power, microwaves transfer electrons into the dot at a much faster rate then the natural electron dwelling time and so, $\Delta \bar{n} = 1$ could be obtained. However, the maximum extrapolated value is 0.78 in our experiment, suggesting feedback actions as described in \ref{One} due to $\it{\Gamma}_{\textup{\tiny{S}}} \ne \textup{0}$. We calculate that $\it{\Gamma}_{\textup{\tiny{S}}} / \it{\Gamma}_{\textup{\tiny{D}}} \sim \textup{0.28}$. Although two electrons could be transfered to the dot, one can tunnel out from the other contact, leaving on average one electron in the dot, so $\Delta \bar{n} \sim 1/2$. When tunneling events are not synchronized, then $\Delta \bar{n} \sim 1$ can be achieved. This explains the switching behavior between the $\Delta \bar{n} \sim 1/2$ and $\Delta \bar{n} \sim 1$ states

\begin{figure}
\begin{center}
\includegraphics[width=85mm, bb=0 0 331 468]{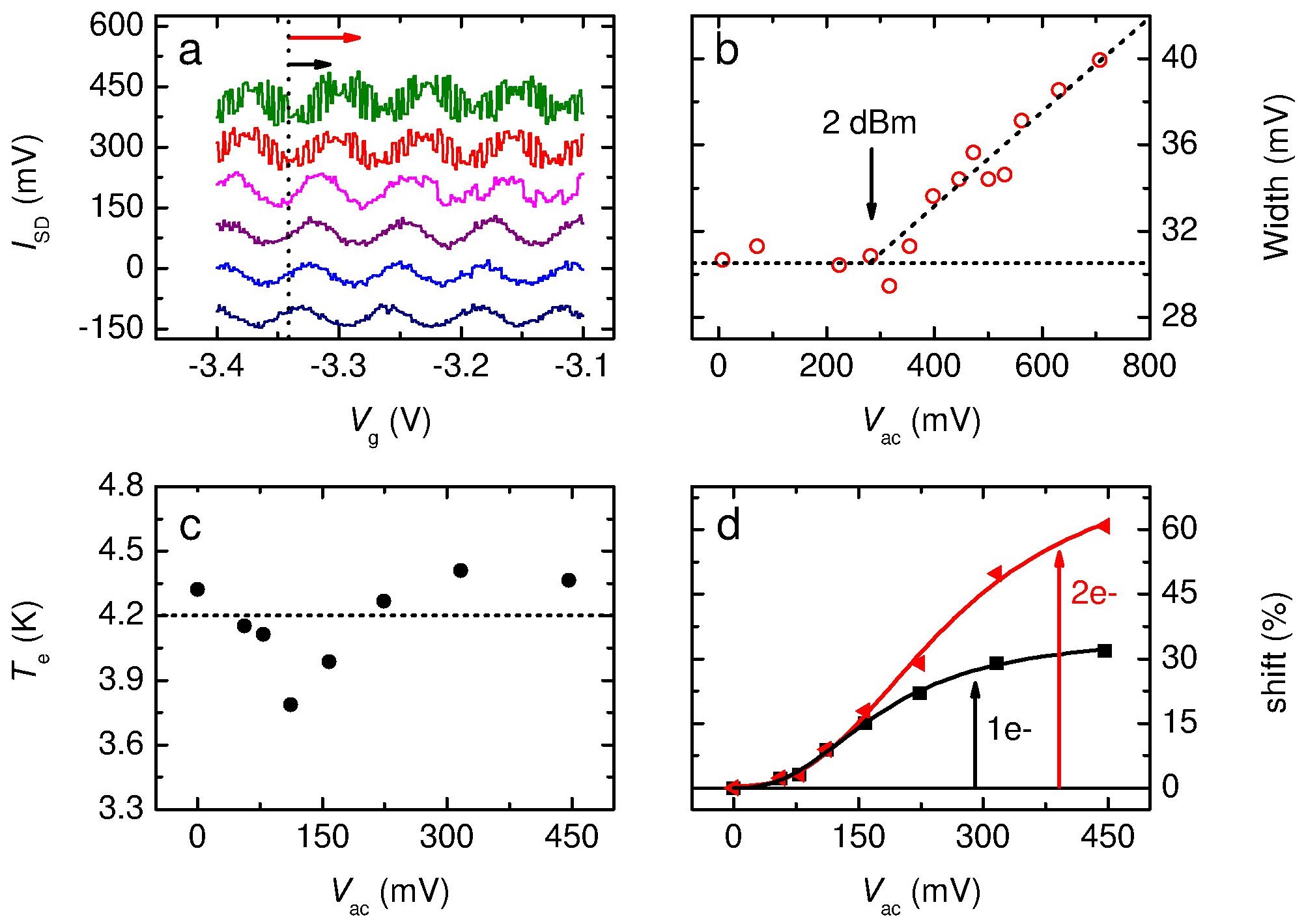}
\end{center}
\caption{\label{fig:figure6} (a) Gate oscillations at different microwave powers, from $P = $6 dBm (top) down to -6 dBm. Last bottom curve is at -135 dBm for reference. Offset was added for clarity. Arrows indicate the value of the shift $\Delta V_{\textup{\tiny{g1}}}$ for the one-electron process (black) and $\Delta V_{\textup{\tiny{g2}}}$ for the two-electron process (red),(b) variation of the Coulomb peak width at half height with the microwave amplitude $V_{\textup{\tiny{ac}}}$. (c) Electron temperature versus $V_{\textup{\tiny{ac}}}$ at $V_{\textup{\tiny{SD}}} = $3 mV. (d) Shifts measured for the one- and two-electron processes as a function of $V_{\textup{\tiny{ac}}}$.}
\end{figure}

\section{Conclusion}

We have shown that under the application of a microwave signal, (i) tunneling of a single or two electrons can be induced while usual conditions for PAT are not met, (ii) a drop-off voltage at the edge of a doped silicon gate does exist and its value can be significantly modified. Whereas the first effect allows a direct manipulation of the electron number in a quantum dot and it is dominant at medium to high power, the second provides at low power an interesting alternative method to gate pulsing, a microwave pulse being equivalent to a voltage pulse on the gate lead. Such results are expected to open the way to qubits manipulation via gigahertz photons in silicon devices. In this approach, the scalability problem to multi-qubit structures does reduce to microwave multiplexing with each individual frequency addressing each individual qubit.

\section{Acknowledgement}

This work was supported by Project for Developing Innovation Systems of the Ministry of Education, Culture, Sports, Science and Technology (MEXT), Japan and by Grants-in-Aid for Scientific Research from MEXT under Grant No. 22246040. T. Kodera would like to acknowledge JST-PRESTO for financial support. T. Ferrus is grateful towards Prof. Sir M. Pepper for discussion and feedbacks on the manuscript, as well as towards Dr Chris Ford, for software and driver development.

\appendix
\section*{Appendix}
\setcounter{section}{1}

\subsection{Single shot measurements}\label{One}

\begin{figure}
\begin{center}
\includegraphics[width=85mm, bb=0 0 331 468]{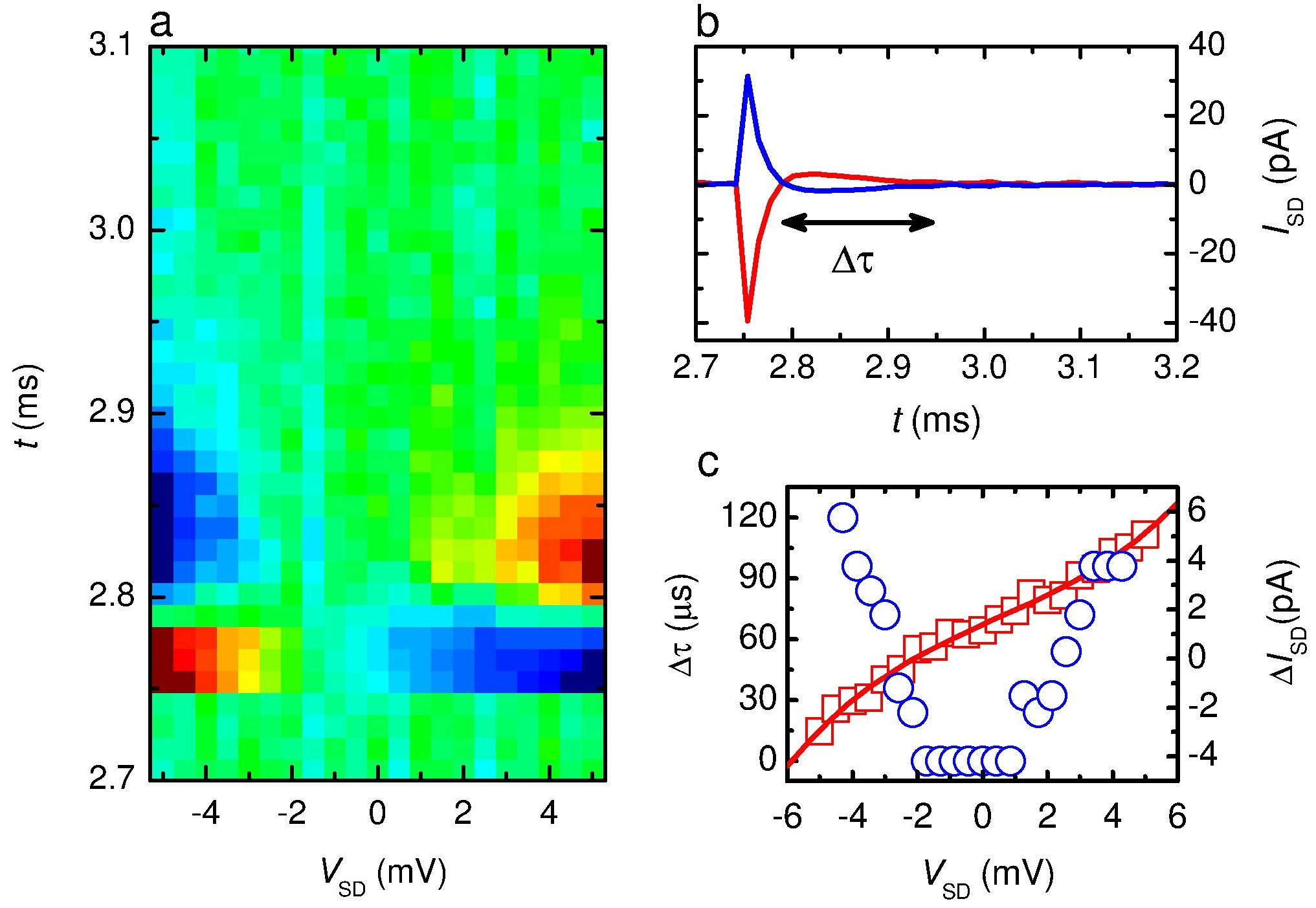}
\end{center}
\caption{\label{fig:figureA1} (a) Single shot measurements at $\nu=5.435$ Ghz and $P=10$ dBm as a function of source-drain bias. (b) Evolution of the current in time, enlightening backaction process for a time $\Delta \tau$. (c) Variation of $\Delta \tau$ and the extremum value of current at $t \sim$ 2.824 ms as a function of $V_{\textup{\tiny{SD}}}$.}
\end{figure}

Unlike continuous wave (continuous irradiation of microwaves) and even continuous pulse measurements (continuous train of short microwave pulses), single shot measurements allow a detailed observation of the electron dynamics inside the dot. Such experiments were carried out by sending a single pulse of microwaves of length 2 $\mu$s, at a frequency of $\nu=5.435$ Ghz and power $P =$10 dBm and, by recording the evolution of the source-drain current $I_{\textup{\tiny{SD}}} \left(t\right)$ with time. This variation is unusual with $I_{\textup{\tiny{SD}}} \left(t\right) < I_{\textup{\tiny{SD}}}\left(0\right) $ for a short time $\Delta \tau$ (Fig.~\ref{fig:figureA1}b). The difference of current before the arrival of the pulse and at the dip increases with both $P$ and $V_{\textup{\tiny{SD}}}$ (Figs.~\ref{fig:figureA1}a and c). This is a clear signature of a backaction from the tunneling electron due to a long $T_1$ lifetime and the presence of electron interaction. When one electron (1) enters the dot under the influence of the microwaves, $\Delta\bar{n}$ gets a non-zero value, shifting the Coulomb diamonds along $V_{\textup{\tiny{g}}}$ before tunneling back to the lead. By increasing $P$, the averaged time spent by the electron (1) in the dot is also increased. Because this effect goes against Coulomb blockade, one of the electrons already present in the quantum dot (2) may tunnel out through the other barrier in order to re-establish the blockade regime and $\Delta\bar{n} = 0$. Because microwaves modify primarily tunnel barriers (where the pair of donors are located), the tunneling rates of the source ($\it{\Gamma}_{\textup{\tiny{S}}}$) and drain barrier ($\it{\Gamma}_{\textup{\tiny{D}}}$) becomes significantly different in values and, there exist a short time $\Delta \tau$ over which the electron (1) has relaxed to the lead whereas the electron (2) has not yet tunnel back to the dot, leading to a decrease of current. 
Figure~\ref{fig:figureA1}c well reflects these effects. By increasing $V_{\textup{\tiny{SD}}}$, the barrier tunnel rates are modified and the dependence of the excess current $\Delta I_{\textup{\tiny{SD}}}$ due to back action is similar to the one of the current through the device. For the same reason, higher is $V_{\textup{\tiny{SD}}}$, longer will be the time spent by the electron (2) outside and longer will be $\Delta \tau$. 

This process is confirmed by measuring the shape of Coulomb diamonds at different time interval. On the arrival of the pulse at $t= 2.75$ ms, diamonds shifts in gate voltage compared with the undisturbed positions, whereas in the dip ($\sim 2.82$ ms), they appear similar to the DC case of Figure 2b, suggesting a superposition of states between an electron outside and inside the dot. Normal diamonds are recovered after 250 $\mu$s ($t \sim 3$ ms).

We also notice that the primary effect is a gate shift rather than a source-drain bias shift, indicating that rectification is not the predominant effect. Because for continuous wave or continuous pulsed measurement the integration time is considerably larger (200 $\times$ 200 $\mu $s = 40 ms), the single shot signal is integrated and the time effect is lost. This is equivalent of summing the two coulomb maps (one with the shift and one without).


\begin{thebibliography}{41}
\bibitem{bleaching} In some circumstances, microwave can affect tunneling in the quantum dot by exciting electrons to higher energy levels (single particle energy levels) and opening new conducting channels to transport. By increasing the microwave power, photon absorption is modified and the SET current initially increases. When the radiative combination is compensated, the current via higher states saturates whereas the one via the low states is reduced to its minimum.
\bibitem{screening} At the device concentration of $N_{\textup{\tiny{D}}} \approx 3\times 10^{19}$ cm$^{-3}$, the average dopant separation is $d_{\textup{\tiny{P-P}}} \sim 1/{N_{\textup{\tiny{D}}}}^{1/3}\sim 3.2$ nm. In this case the Thomas-Fermi approximation is valid and the screening length is ${\lambda}_{\textup{\tiny{S}}} = {\pi}^{2/3}\sqrt{\frac{\epsilon {\hbar}^2}{e^2 m \left(3 N_{\textup{\tiny{D}}} \right)^{1/3}}} \sim 16$ nm.
\bibitem{tunnel rate} Considering that microwaves act as a small perturbation (addition of an oscillatory term of amplitude $V_{\textup{\tiny{ac}}}$ in the Hamiltonian), the wavefunctions $\mid {\psi}_{i\textup{\tiny{0}}} >$ and $\mid {\psi}_{j\textup{\tiny{0}}} >$ associated with each donors on both side are modified, so that $\mid {\psi}_{i} > = \mid {\psi}_{i\textup{\tiny{0}}} > + \alpha V_{\textup{\tiny{ac}}} \mid {\psi}_{j\textup{\tiny{0}}} >$. The tunneling probability is then ${\mathcal{P}}_{ij} \propto {V_{\textup{\tiny{ac}}}}^2$ and so the tunneling current and the tunneling rate are proportional to the effective power at the device, on the first approximation.
\end{thebibliography}
\end{document}